\documentclass[twocolumn,showpacs,preprintnumbers,amsmath,amssymb,pre,aps]{revtex4-1}
\pdfoutput=1
\usepackage{graphicx,color}% Include figure files
\usepackage{url,float}
\usepackage[percent]{overpic}
\usepackage{rotating} 
\usepackage{minibox}

\newlength{\defbaselineskip}
\newcommand{\setlinespacing}[1]%
           {\setlength{\baselineskip}{#1 \defbaselineskip}}

\begin{document}

% \title{Extracting Unbiased Estimates of Kinetic Confinement  Parameters  from Short Single-Molecule Trajectories Containing  Measurement Noise}
\title{Robust Hypothesis Tests for Detecting  Statistical Evidence of 2D and 3D Interactions in Single-Molecule Measurements}

% \textbf{Technical Report for the Moerner and Scott Labs:  \hfill  Please do not distribute. \\ }

\begin{abstract} 
 A variety of experimental techniques have
improved the 2D and 3D spatial resolution that can be extracted from \emph{in vivo} single-molecule measurements.  
This enables researchers to quantitatively infer the magnitude and directionality of forces experienced by 
biomolecules in their native cellular environments. 
Situations where such forces are biologically relevant range from mitosis to directed transport of protein cargo along  cytoskeletal structures. 
Models commonly applied to quantify single-molecule dynamics assume that effective forces and velocity 
in the $x,y$ (or $x,y,z$) directions are statistically independent, but this  assumption is physically unrealistic
in many situations.   
We present a hypothesis testing approach capable of determining if there is evidence of 
statistical dependence  between  positional coordinates  in experimentally measured trajectories; 
if the hypothesis of independence between spatial coordinates is rejected, then a new model accounting 
for 2D (3D) interactions should be considered to more faithfully represent the underlying experimental kinetics. 
The technique is robust in the sense that 2D (3D) interactions can be detected via statistical hypothesis testing
 even if there is substantial inconsistency between the physical particle's actual noise sources and the 
 simplified model's assumed noise structure. 
 For example, 2D (3D) interactions can be reliably detected even if the  researcher assumes normal diffusion, but the experimental data experiences  ``anomalous diffusion" and/or is subjected to a measurement noise characterized by a 
  distribution differing from that  assumed by the fitted model.  The approach is 
 demonstrated on control simulations and on experimental data 
 (IFT88 directed transport in the primary cilium).   

  \end{abstract}

\author{\small Christopher P. Calderon $^\dagger$}
\email{chris.calderon@numerica.us}
\author{\small Lucien E. Weiss $^\ddagger$}
\author{\small W. E. Moerner $^\ddagger$}
%APS format

% \affiliation{%
% $^\dagger$  \small Numerica Corporation, 4850 Hahns Peak Drive, Loveland, Colorado, 80538.
% }
\affiliation{%
$^\dagger$  \small Numerica Corporation,  Loveland, Colorado, 80538
}
\affiliation{%
$^\ddagger$  \small Department of Chemistry, Stanford University, Stanford, California, 94305.
}

%SPIE format
% \authorinfo{Author E-mail: chris.calderon@numerica.us
% \center{\small{{\it }}}}

%APS format
% PACS: 02.50.-r  87.10.Mn

\date{ \today}

\maketitle

%%%%%%%%%%%%%%%%%%%%%%%%%%%%%%%%%%%%%%%%%%%%%%%%%%
\section{Introduction} Several advances in optical microscopy have increased the spatial and temporal resolution
that can be extracted from single particle tracking (SPT) experiments. For example,
the techniques used in Refs. \cite{Arhel2006,Golding2006a,Lessard2007,Lange2008,manley2008,Thompson2010,Weigel2011,Ram2012,Ye2013} enable 
researchers to experimentally monitor the 2D (or 3D) kinetics of a fluorescently tagged biomolecule  
%either \emph{in vitro} or 
\emph{in vivo}.  However, 
 commonly used SPT analysis techniques  do not explicitly model 2D or 3D spatial interactions 
 despite the fact that spatially dependent force interactions
 or anisotropic diffusion can be important in many biological processes 
 including intracellular trafficking \cite{Ram2012}, diffusion in confined environments \cite{Ye2013}, molecular motor induced transport \cite{Thompson2010}, chromatin dynamics \cite{Verdaasdonk2013},
 and mitosis \cite{Stephens2013}.

 Recent articles related to SPT analysis methods have studied 2D/3D estimation \cite{Masson2009,Voisinne2010,Calderon2013b} 
 and goodness-of-fit (GoF) testing \cite{Calderon2013b}  (i.e., checking the consistency of  model assumptions 
 against individual experimental trajectories containing thermal and measurement noise).
 GoF tests are helpful in checking   the various statistical 
 assumptions implied by an assumed model, 
 but GoF tests  looking for non-specific model imperfections are not always desirable in SPT data
 analysis. For example, suppose a researcher
 observes a 2D time series generated by an SPT experiment
 (positions of the tagged particle are denoted by $x$ and $y$) and subsequently uses this to estimate the parameters 
 of a  standard  diffusion, but the underlying data is generated by a process 
 with anomalous diffusion  (i.e.,   $H\ne \frac{1}{2}$  where $H$ is the Hurst exponent \cite{Kou2004,kou_08,Magdziarz2010,Weber2012,Kepten2013,Meroz2013}).   
 If a  GoF test is used and the model is rejected, the GoF test  would not directly  reveal 
 the specific reason for rejection \cite{Bickel2006,Calderon2013b}.
 In this example, the  incorrect thermal noise model would ultimately lead to rejection by a consistent GoF test
 regardless of whether or not unmodeled statistical dependence exists between the $x$ and $y$ coordinates. 

In situations where a researcher is solely interested in determining if statistically significant evidence 
of 2D or 3D interactions 
exists in the experimental data, 
 he/she may want to  specifically test for any dependence between $x,y$ (or $x,y,z$) 
 measurements.  
 It would be desirable to have such a hypothesis test be robust to 
 questionable modeling assumptions 
 which can be challenging to test in short  SPT trajectories 
 containing both thermal fluctuations and noise induced by the measurement apparatus. For example, the anomalous vs. standard diffusion 
 question can be hard to resolve using a single trajectory when one only has access to a 
 time series spanning $\approx 1-5 \ s$ and
  measurement noise  is significant relative to 
 thermal fluctuations \cite{Weber2012,Kepten2013,Meroz2013} (this noise is sometimes referred to as ``localization noise'' \cite{Berglund2010} in the SPT literature).  
 Robustness against 
  questionable localization noise assumptions is desirable 
  since this noise is difficult to accurately 
 quantify in many experiments \cite{Ait-Sahalia2010,Berglund2010,Michalet2012,Calderon2013b} despite its potential heavy influence on parameter estimation, hypothesis testing, and other model diagnostics \cite{TSRV,Ait-Sahalia2012,Calderon2013}. 
 In addition, methods not requiring ensemble averaging  (i.e., those capable of carrying out tests with a single trajectory) 
 are of interest since 
 the effective dynamics experienced \emph{in vivo} can be heterogeneous due to varying local micro-environments    \cite{Calderon2013b,Kepten2013,Meroz2013}.

 In this work, we present such a hypothesis testing technique.  This is accomplished by combining 
 techniques discussed in Refs. \cite{Calderon2013b} and \cite{Duchesne2012}.    
Practical utility of the approach  is 
demonstrated through both simulation and experiments.  
Although we focus on situations motivated by SPT experiments, the techniques are applicable 
to other single-molecule experiments involving  2D (or 3D) time series measurements.  
%(time spacing need not be uniform). 

\section{Methods} 
\subsection{Data Generating Process (DGP) Used in Simulation Studies}
For the simulation cases studied, a stochastic differential equation (SDE) model of the form:
\begin{align}
\label{eq:SDE}
d\vec{r}_t=& \frac{\sigma\sigma^{\mathrm{T}}}{k_BT}F\vec{r}_tdt+\sqrt{2} \sigma d\vec{B}_t^H &\\
\label{eq:meas}
 \vec{\psi}_{k}=& \vec{r}_{k}+\vec{\epsilon}_k, %\ \ \ \vec{\epsilon}_i  \sim \mathcal{N}(0,R),
\end{align}
is used as the data generating process (DGP).  In the expression above, $\vec{r}_t \in \mathbb{R}^d$ denotes the position of the tagged particle 
(e.g., if $d=2$ then $r=(x,y)^{\mathrm{T}}$ where $^{\mathrm{T}}$ denotes the transpose operation); $k_BT$ is the product of Boltzmann's constant and temperature;
the product of $F\in \mathbb{R}^{d\times d}$  
and $\vec{r}_t$ determines the instantaneous force experienced by the particle; $\vec{B}_t^H$ is a fractional 
Brownian motion (fBm)  \cite{kou_08,Kepten2013,Meroz2013} \footnote{
The drift function in the $H=\frac{1}{2}$ case is motivated by the overdamped Langevin equation; 
for $H\ne\frac{1}{2}$,  a memory kernel is not used; however our interest is 
in robustly detecting statistical dependence between components regardless of fluctuation dissipation constraints.}; 
and $\vec{\psi}_{k}$ is the measurement/observation vector taken at time $t_k$. 
Note that the position is not directly observed due experimental artifacts like localization noise \cite{Berglund2010,Weber2012}; the latter is modeled a mean zero 
Gaussian noise
$\vec{\epsilon}_k$ having covariance $R$.  Both $\sigma$ and $R$ are diagonal matrices $\in \mathbb{R}^{d\times d}$.
% All vectors and matrices have  dimension consistent with $\vec{r}_t$.  
To specify the model
parameters, we report possible nonzero elements of  matrices; for example, if $d=2$, then $F$ will have four parameters 
($F_{11}$,$F_{21}$,$F_{12}$,$F_{22}$) and 
$R$ will have two ($R_{11}$,$R_{22}$) where subscripts  denote  rows/columns. 

\subsection{Model and Hypothesis Testing Procedure Outline}
To model observations, the multivariate SDEs (with $H=\frac{1}{2}$) presented  in Ref. \cite{Calderon2013b} are utilized. % (all models assume $H=\frac{1}{2}$).  
This assumed model structure allows
one to utilize exact Kalman filter likelihood equations for both estimation and inference.  
After  applying maximum likelihood estimation (MLE), one can use  %(which may or may not permit 2D/3D interactions)
the estimated MLE parameter, the model's implied conditional distribution, and the time series 
observations ($\{\psi_{k}\}_{k=1}^N$)
to generate the normalized
innovation sequence $\{\vec{e}_{k}\}_{k=1}^N$ \cite{Calderon2013b}. In what follows (using notation from Ref. \cite{Duchesne2012}), components of the estimated normalized innovation vector, $\vec{e}_{k}$, are denoted by $e_{jk}$ where the first subscript indexes the position coordinate and the second subscript indexes time.

 In Ref.  \cite{Calderon2013b}, the components of different coordinates were combined into a single time series, e.g. a one dimension vector of the form $(e_{11},e_{21},e_{31},e_{12},e_{22}, \ldots e_{3N})$ was constructed for each MLE; subsequently, the  GoF tests
reported in Ref. \cite{hong} were utilized.  The problem with this approach is that temporal and spatial measurement information is aggregated together.  This can result in a loss of power in detecting model imperfections caused by time dependent interactions between different spatial coordinates.  Furthermore, Hong and Li's tests are GoF tests seeking to detect any model imperfection (some modeling errors may not be of scientific interest).
% (i.e., the GoF test does not specifically seek to detect statistical dependence between different spatial coordinates).

%generalized residuals.  aggregate information unnecessarily  can keep strict time and space order info.
%
%aggregation can lose power.
%
%small outliers can adversely affect results.  CVM motivation.
%
%method in \cite{Duchesne2012}  can be used with a variety of residuals (generalized).  
%applicable to linear and nonlinear time series.  as in \cite{Calderon2013}, we study the goodness of fit problem by finding the MLE and using this to transform the raw data into a normalized innovation sequence.  1D models do not allow coupling, so 

The methods of Duchesne \emph{et al.} focus on tests that depend on the 
 empirical rank statistics of estimated residuals \cite{Duchesne2012} (in our situation, the residuals are the normalized innovations \cite{Calderon2013}).  
 Duchesne \emph{et al.}'s approach utilizes a multivariate time series analysis
 approach applicable to two or more components (technical complications associated with testing time series vectors in $\mathbb{R}^3$ are readily handled \cite{Duchesne2012}).  The multivariate approach respects the natural time and space ordering of the raw observations (in contrast the GoF approach used in Ref. \cite{Calderon2013}  where a $d=3$ vector was collapsed to a $d=1$ vector as illustrated in the
 previous paragraph).  
 %their tests also check specifically for statistical
 %dependence between two or more components.  
%
 To achieve this, 
Duchesne \emph{et al.} developed a technique that uses the so-called M\"{o}bius 
transformation to map a collection of empirical distributions to a collection of asymptotically independent Gaussian random variables 
under the null hypothesis of independence.  
 In this study, we compute $\mathcal{H}_N$ (the generalized cross correlation statistic discussed in Ref. \cite{Duchesne2012}) to test for dependence.  We also study the related  Cramer-von Mises   
 test statistic, $\mathcal{W}_N$ \cite{Duchesne2012}. 
 The Appendix  summarizes and discusses the equations 
  originally presented in  Ref. \cite{Duchesne2012} that we utilize in this work. 
   Note that we study multiple test statistics for both GoF and independence testing since  selecting the 
 ``best'' test statistic in finite sample sizes where spatial and/or temporal dependence exists in the observational data can be difficult
 \cite{Bickel2006}.

\setcounter{figure}{0}
\begin{figure}[htb]
\center
\centering

\begin{minipage}[b]{.95\linewidth}
\def\pw{1}

% \begin{overpic}[width=\pw\textwidth]{Figure1.png}
% % \put(15,72){\Large \color{black}(d)}
% % \put(35,-5){{\makebox[1.\width]{\large 1D Model} \hfill }}
% \put(5,-3){\mbox{\parbox[b]{0.5\textwidth}{\normalsize GoF Tests \hfill } }}
% \put(50,-3){\mbox{\parbox[b]{0.5\textwidth}{\normalsize Independence Tests \hfill } }}
% \end{overpic}

\begin{overpic}[width=\pw\textwidth]{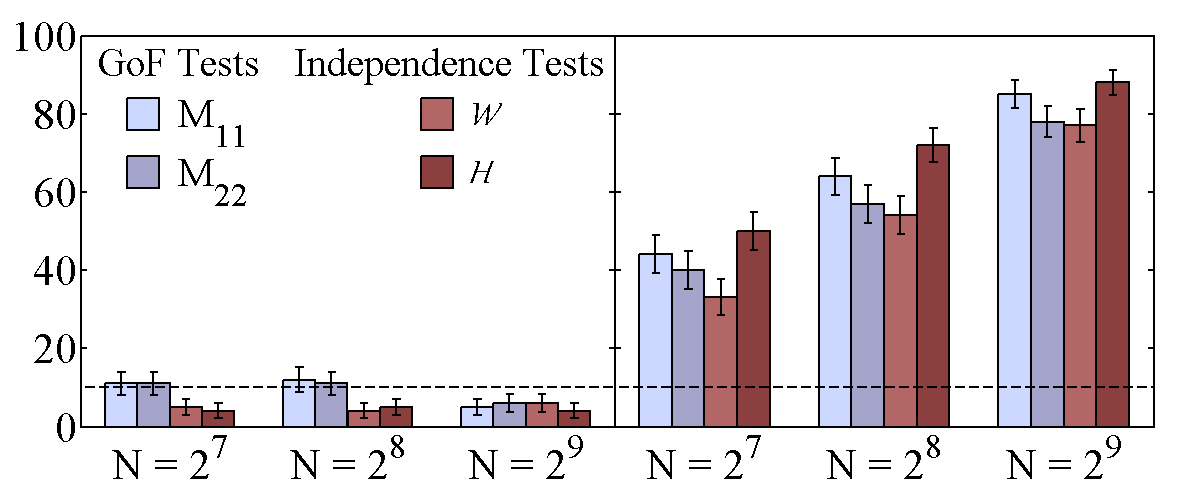}
% \put(15,72){\Large \color{black}(d)}
% \put(35,-5){{\makebox[1.\width]{\large 1D Model} \hfill }}
\put(-4,7){\rotatebox{90} {\makebox[1.\width]{\large $\%$ Rejected} \hfill }}
\put(5,-5){\mbox{\parbox[b]{0.5\textwidth}{\normalsize Exact 2D Model \hfill } }}
\put(50,-5){\mbox{\parbox[b]{0.5\textwidth}{\normalsize Incorrect 1D Model  \hfill } }}
\end{overpic}
  \end{minipage}

\centerline{\footnotesize }
\caption{%\footnotesize 
\footnotesize 
% Percent rejections obtained with  test statistics applied to two models (vertical lines denote error bars). Dashed horizontal line displays expected percent rejection under null and infinite sample sizes. ``2D'' (``1D'') denotes results obtained using the 2D SDE (1D SDE) models discussed in main text; the 2D SDE model matches the DGP precisely here. 
(Color online).  Percent rejections obtained with four test statistics applied to two models (vertical lines denote error bars). Dashed horizontal line displays expected percent rejection under null and infinite sample sizes. (Left) 2D model with $H=\frac{1}{2}$ and  correlation (induced by $F$) matching the DGP. (Right) Incorrect 1D model neglecting statistical dependence.   
}
\label{fig:1}
\end{figure}

% \setcounter{figure}{0}
% \begin{figure}[htb]
% \center
% \centering

% \begin{minipage}[b]{.95\linewidth}
% \def\pw{1}

% % \begin{overpic}[width=\pw\textwidth]{Figure1.png}
% % % \put(15,72){\Large \color{black}(d)}
% % % \put(35,-5){{\makebox[1.\width]{\large 1D Model} \hfill }}
% % \put(5,-3){\mbox{\parbox[b]{0.5\textwidth}{\normalsize GoF Tests \hfill } }}
% % \put(50,-3){\mbox{\parbox[b]{0.5\textwidth}{\normalsize Independence Tests \hfill } }}
% % \end{overpic}

% \begin{overpic}[width=\pw\textwidth]{figure1_alt.png}
% % \put(15,72){\Large \color{black}(d)}
% % \put(35,-5){{\makebox[1.\width]{\large 1D Model} \hfill }}
% \put(5,-3){\mbox{\parbox[b]{0.5\textwidth}{\normalsize GoF Tests \hfill } }}
% \put(50,-3){\mbox{\parbox[b]{0.5\textwidth}{\normalsize Independence Tests \hfill } }}
% \end{overpic}
%   \end{minipage}

% \centerline{\footnotesize }
% \caption{%\footnotesize 
% \footnotesize 
% \textbf{[Direct ``Window Print"]} Percent rejections obtained with  test statistics applied to two models (vertical lines denote error bars). Dashed horizontal line displays expected percent rejection under null and infinite sample sizes. ``2D'' (``1D'') denotes results obtained using the 2D SDE (1D SDE) models discussed in main text; the 2D SDE model matches the DGP precisely here.   
% }
% \label{fig:1b}
% \end{figure}

\subsection{Simulation Parameters}
In all cases studied, the trajectory observations are $\{\vec{\psi}_{k}\}_{k=1}^N$ where 
$10$ $ms$ separate observations and the system temperature is $310$ K.  For the simulations, 400 Monte Carlo trajectories were used to estimate MLEs and test statistics.
The SDE parameters are: 
$F=(-4,-.2,-1,-.1)$ [$\frac{pN}{\mu m}$],
$\sigma=(.2,.3)$ [$\frac{\mu m}{s^{1/2}}$],
$R=(30,20)$ [$nm$] for the 2D simulations and  
$F=(-4,-.2,0,-1,-.1,.1,0,0-.1)$ [$\frac{pN}{\mu m}$],
$\sigma=(.2,.3,.3)$ [$\frac{\mu m}{s^{1/2}}$], 
$R=(30,20,20)$ [$nm$] for 3D.

\subsection{Experimental Details}
IFT88-eYFP tracking experiments in live cultured cells \cite{Tran2008} were performed using an Olympus IX-71 inverted 
microscope with a 100x, 1.4 NA oil objective (UPLSAPO, Olympus). Cell culture was performed as previously 
described \cite{Ott2012}. Two days prior to experiments, cells were detached (Invitrogen, TripLE Express) 
from cell culture dishes (Thermo Scientific, tissue culture treated 10 cm dish) and replated onto cell-culture filters 
(Corning, Costar Transwell 0.4 $\mu m$ Polyester Membrane 6.5 mm Inserts) in low-serum DMEM media 
(Thermo, 0.5\% Hyclone Fetal Bovine Serum; Invitrogen, Gibco phenol red-free Dulbecco's Modified Eagle Medium) to induce cilia 
growth. At the start of the experiment, filters were placed onto \#1 glass coverslips 
(Fisher) and contained in a humidity and temperature controlled stage-top incubator (Tokai-Hit, ONICS). 
The sample was illuminated with circularly polarized, 514 nm pumping light (Coherent Sapphire, 50 mW). 
Fluorescence collected through the objective was filtered with a dichroic beamsplitter (Semrock, FF425/532/656-Di01) 
and a bandpass filter (Semrock, FF01-578/105-25) before being imaged in 10 ms frames on an electron multiplied charge-coupled device (Andor Technology, iXon Ultra DU897). 
Single-molecule trajectories were gathered from the subsequent movie using custom Matlab image-analysis 
software that fit a 2D symmetric Gaussian with a constant offset to a small region of the image. Bad fits were removed if the fitted parameters were determined to be inconsistent with those of a single YFP molecule.

\section{Results} 
The results are divided into two sections.  The first presents the (unrealistic) situation where one of the assumed stochastic
models precisely matches the DGP (i.e., all noise distributions and spatial/temporal dependencies are exactly known). The second set of results shows the more practical situation where the DGP has features
not accounted for in the fitted model.  GoF procedures and hypothesis tests aiming to detect unmodeled statistical dependence are studied in both
situations.

\subsection{Control Simulations: \newline \emph{The DGP Matches an Assumed Model}}
\label{sec:control}
In Fig. \ref{fig:1}, a 2D SDE with $H=\frac{1}{2}$ and correlation  between $x$ and $y$  serves as the DGP. Here and in what follows,
``correlated'' refers to dependence induced by non-zero off-diagonal terms in $F$  and ``uncorrelated'' refers to situations where
$F$ is a diagonal matrix. 
Two models are applied to each simulated trajectory.  One estimated model, referred to as ``2D SDE'' has the correct parametric structure;  the other model constrains the off-diagonal entries of $F$ to be zero resulting in an uncorrelated model 
(this model is also referred to as the ``1D SDE'' since  one estimate model parameters without jointly observing  $x$ and $y$).  For each trajectory, the MLE parameter  is computed (see Ref. \cite{Calderon2013b})
along with four tests statistic $\mathcal{W}_N$, $\mathcal{H}_N$, $M(1,1)$, and $M(2,2)$;  the first two test statistics aim to robustly detect unmodeled statistical dependence \cite{Duchesne2012} 
% (the null hypothesis is independence between the computed normalized innovations)  
and the latter two check GoF \cite{hong};
% (the null hypothesis is that the fitted parametric model matches the exact distribution of the data generating process); 
these four test statistics are computed for each individual trajectory.  The fraction of rejections are plotted for various $N$ (nominal expected rejection rate under the null hypotheses shown by dashed horizontal line). 
Table \ref{tab:1} contains analogous results, except a 3D SDE with $H=\frac{1}{2}$  generates data.  These results demonstrate the Type I error rate and power under situations and sample sizes relevant to various SPT applications.  Note that $\mathcal{H}_N$ improves
 power in detecting 2D/3D interactions in the cases studied 
(comparable power improvement also occurs if the 3D simulations in Ref. \cite{Calderon2013b} are studied).

\begin{table} [htb] 
 \center 
 \caption{\label{tab:} \footnotesize Hypothesis Testing Results with 3D Data Generating Process.  Fraction rejected using nominal Type I rate $=10\%$. For the (incorrect) 1D  model, off-diagonal entries of $F$ are set to $0$. } 

%  \begin{tabular}{|l|*{4}{c|}}\hline

%  \multicolumn{5}{|c|}{ \textbf{3D Model (exact model)}}   \\ \hline %put multliline in later \\ 
% &  $\mathcal{W}$ & $\mathcal{H}$ & $M_{11}$ &  $M_{22}$ \\  \hline 
% $N=2^8$ & 10.25 & 9.25 & 5.50 & 7.00 \\  \hline 
% $N=2^9$  & 9.00 & 9.50 & 6.25 & 6.75 \\  \hline
%    \multicolumn{5}{|c|} { \textbf{1D Model (misspecified model)}}  \\  \hline %put multliline in later
%   & $\mathcal{W}$ & $\mathcal{H}$ & $M_{11}$ &  $M_{22}$ \\  \hline   
% $N=2^8$  & 25.25 & 32.25 & 21.50 & 16.25 \\  \hline 
% $N=2^9$  & 43.00 & 58.00 & 40.25 & 33.75 \\  \hline 
% \end{tabular}

 \begin{tabular}{|l|*{9}{c|}}\hline

%  \multicolumn{5}{|c|}{\textbf{3D Model (exact model)}}   \\ \hline %put multliline in later \\
% &  $\mathcal{W}$ & $\mathcal{H}$ & $M_{11}$ &  $M_{22}$ \\  \hline
% $N=2^8$ & 10.3 & 9.3 & 5.5 & 7.0 \\  \hline
% $N=2^9$  & 9.0 & 9.5 & 6.3 & 6.8 \\  \hline
%    \multicolumn{5}{|c|} {\textbf{1D Model (model mismatch)}}  \\  \hline %put multliline in later
%   & $\mathcal{W}$ & $\mathcal{H}$ & $M_{11}$ &  $M_{22}$ \\  \hline
% $N=2^8$  & 25.3 & 32.3 & 21.5 & 16.3 \\  \hline
% $N=2^9$  & 43.0 & 58.0 & 40.3 & 33.8 \\  \hline

& \multicolumn{4}{c|}{\textbf{Exact 3D Model}} & & \multicolumn{4}{c|}{\textbf{1D Model}}   \\  %put multliline in later \\
 %\multicolumn{1}{|c|}{$N$}  & \multicolumn{4}{c|}{\textbf{(exact DGP)}} & & \multicolumn{4}{|c|}{\textbf{(incorrect)}}   \\ \hline %put multliline in later \\
% &  $\mathcal{W}$ & $\mathcal{H}$ & $M_{11}$ &  $M_{22}$ & & $\mathcal{W}$ & $\mathcal{H}$ & $M_{11}$ &  $M_{22}$ \\  \hline
\multicolumn{1}{|c|}{$N$}  & \multicolumn{1}{c}{$\mathcal{W}$ } & \multicolumn{1}{c}{$\mathcal{H}$ } & \multicolumn{1}{c}{$M_{11}$ } & \multicolumn{1}{c|}{$M_{22}$ } & & \multicolumn{1}{c}{$\mathcal{W}$ } & \multicolumn{1}{c}{$\mathcal{H}$ } & \multicolumn{1}{c}{$M_{11}$ } & \multicolumn{1}{c|}{$M_{22}$ }    \\ 
$2^8$ & 10.3 & 9.3 & 5.5 & 7.0  & & 25.3 & 32.3 & 21.5 & 16.3 \\ \hline
$2^9$  & 9.0 & 9.5 & 6.3 & 6.8  & & 43.0 & 58.0 & 40.3 & 33.8 \\  \hline
% $N=2^8$ & 10.3 & 9.3 & 5.5 & 7.0 \\  \hline
% $N=2^9$  & 9.0 & 9.5 & 6.3 & 6.8 \\  \hline
   
%    \\  \hline
% $N=2^8$  & 25.3 & 32.3 & 21.5 & 16.3 \\  \hline
% $N=2^9$  & 43.0 & 58.0 & 40.3 & 33.8 \\  \hline

\end{tabular}

\label{tab:1}
 \end{table} 

\subsection{Robustness Studies: \newline \emph{The DGP Does Not Match Assumed Model}}

\begin{figure}[htb]
\center
\centering

\begin{minipage}[b]{.95\linewidth}
\def\pw{1}

\begin{overpic}[width=\pw\textwidth]{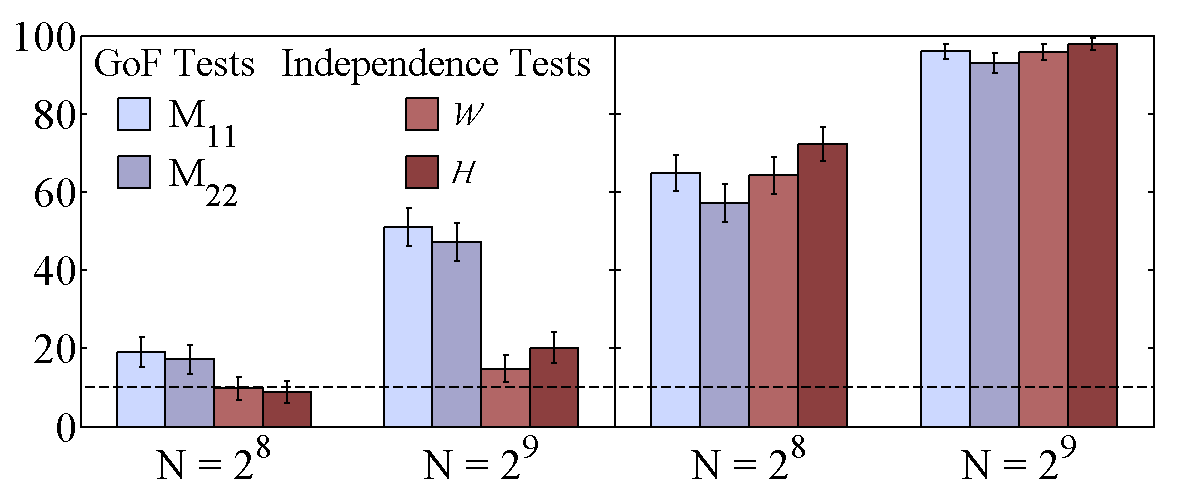}
% \put(15,72){\Large \color{black}(d)}
% \put(35,-5){{\makebox[1.\width]{\large 1D Model} \hfill }}
\put(-4,7){\rotatebox{90} {\makebox[1.\width]{\large $\%$ Rejected} \hfill }}
\put(5,-5){\mbox{\parbox[b]{0.5\textwidth}{\normalsize Incorrect 2D Model  \hfill } }}
\put(50,-5){\mbox{\parbox[b]{0.5\textwidth}{\normalsize Incorrect 1D Model  \hfill } }}
\end{overpic}
  \end{minipage}

\centerline{\footnotesize }
\caption{%\footnotesize 
\footnotesize 
% (Color online) Correlated fBm ($H=0.74$) example.  Similar to Fig. \ref{fig:1} except DGP uses $H=0.74$ noise (models use $H=\frac{1}{2}$, so no model matches all distributional assumptions of DGP).
%  
(Color online) DGP uses correlated anomalous diffusion,  with $H = .74$.  Both models incorrectly assume normal ($H = \frac{1}{2}$) diffusion. 2D model allows for correlation; 1D model  sets correlation to zero.
}
\label{fig:3}
\end{figure}

Next we turn to situations where the DGP does not match the assumed model.  The previously studied 2D DGP is modified to have $H=0.74$ \cite{kou_08}; note that $\sigma$ is multiplied by 250 to keep trajectories comparable, the remaining SDE parameters are  identical.  
The $H>\frac{1}{2}$ case was studied because measurement noise can significantly
complicate  detection of artifacts induced by thermal noise with $H<\frac{1}{2}$ in the ``high frequency'' and relatively small $N$  sampling regime studied \cite{Weber2012,Ait-Sahalia2012}.
Fig. \ref{fig:3} demonstrates that the  GoF tests are able to detect the long range correlation in noise (and the unmodeled spatial correlation).  The dependence test for $\mathcal{W}_N$ and $\mathcal{H}_N$ are just above the nominal null threshold in the 2D SDE model (unmodeled statistical dependence between $x$ and $y$ is again readily detected in the 1D SDE model).  Although the (incorrect) 2D SDE model with $H=\frac{1}{2}$ is able to remove 
much of the linear correlation, statistical dependence between the residuals ($e_{1k}$ and $e_{2k}$) 
 still exists due to the actual  DGP having $H=0.74$; as more data is obtained, it becomes easier to detect this 
 dependence, hence rejection rates are above the nominal null level even in the 2D model.

\begin{figure}[htb]
\center
\centering

\begin{minipage}[b]{.95\linewidth}
\def\pw{.8}

\begin{overpic}[width=\pw\textwidth]{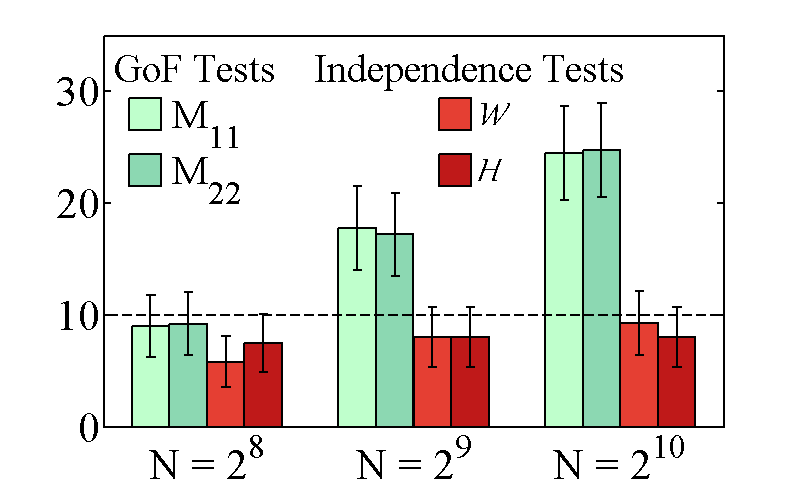}
% \put(15,72){\Large \color{black}(d)}
% \put(35,-5){{\makebox[1.\width]{\large 1D Model} \hfill }}
\put(-3,17){\rotatebox{90} {\makebox[1.\width]{\large $\%$ Rejected} \hfill }}
% \put(-8,-5){\mbox{\parbox[b]{1.\textwidth}{\normalsize  $H=\frac{1}{2}$ Model  (DGP 1D Model with $H=0.74$)   \hfill } }}
\put(-8,-5){\mbox{\parbox[b]{1.\textwidth}{\normalsize  Incorrect 2D Model     \hfill } }}
% \put(50,-3){\mbox{\parbox[b]{0.5\textwidth}{\normalsize 1D Model [$H=\frac{1}{2}$] \hfill } }}
\end{overpic}
  \end{minipage}

\centerline{\footnotesize }
\caption{%\footnotesize 
\footnotesize 
(Color online) DGP uses uncorrelated anomalous diffusion,  with $H = .74$. New coloring scheme emphasizes that no statistical dependence exists between $x$ and $y$ in the DGP.  Incorrect 2D model allows for potential correlation (but uses incorrect $H$).
% 2D SDE with $H=\frac{1}{2}$ model allowing for correlation fit to data and tested, but  
%  process with $H=0.74$ and no dependence (off-diagonal $F$ entries set to zero) was used by DGP.   
 %
 The GoF Tests detect signatures of $H=0.74$ whereas the Independence Tests (correctly) achieve the expected rejection under  null (null of the latter tests is consistent with the DGP). 
}
\label{fig:4}
\end{figure}

\begin{figure}[htb]
\center
\centering

\begin{minipage}[b]{.95\linewidth}
\def\pw{.8}

\begin{overpic}[width=\pw\textwidth]{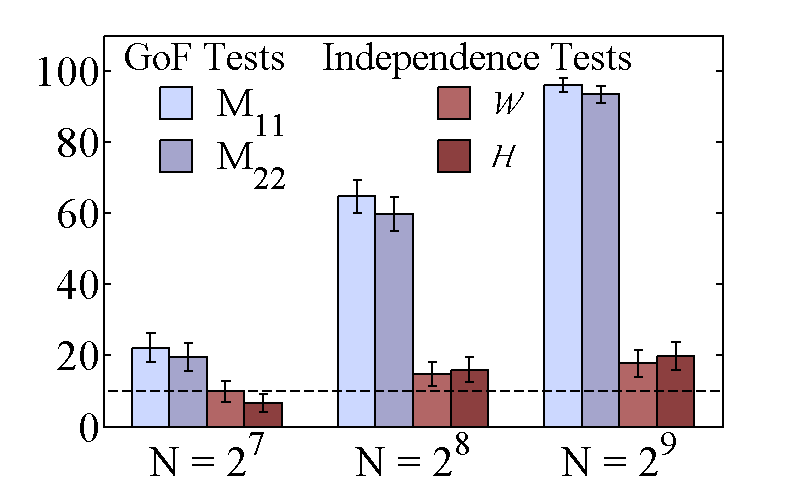}
% \put(15,72){\Large \color{black}(d)}
% \put(35,-5){{\makebox[1.\width]{\large 1D Model} \hfill }}
\put(-3,17){\rotatebox{90} {\makebox[1.\width]{\large $\%$ Rejected} \hfill }}
% \put(-8,-5){\mbox{\parbox[b]{1.\textwidth}{\normalsize  $H=\frac{1}{2}$ Model  (DGP 1D Model with $H=0.74$)   \hfill } }}
\put(-8,-5){\mbox{\parbox[b]{1.\textwidth}{\normalsize  Incorrect $R=0$ Model     \hfill } }}
% \put(50,-3){\mbox{\parbox[b]{0.5\textwidth}{\normalsize 1D Model [$H=\frac{1}{2}$] \hfill } }}
\end{overpic}
  \end{minipage}

\centerline{\footnotesize }
\caption{%\footnotesize 
\footnotesize 
(Color online) DGP uses correlated standard diffusion with measurement noise.  The 2D model allows 
for potential correlation in the drift, but incorrectly assumes no measurement noise is present (i.e., the model 
sets $R=0$ without estimating this parameter from observed data).
% 2D SDE with $H=\frac{1}{2}$ model allowing for correlation fit to data and tested, but  
%  process with $H=0.74$ and no dependence (off-diagonal $F$ entries set to zero) was used by DGP.   
 %
 The GoF Tests readily detect signatures of $R>0$ in the data. The Independence Tests are relatively robust to
 the modeling error, but the innovations are not statistically independent due to measurement noise modeling errors (see discussion in main text). 
}
\label{fig:4b}
\end{figure}

To demonstrate that the method of Ref. \cite{Duchesne2012} is  robust to an incorrectly assumed model
where no $x$/$y$ dependence exists, the previous DGP is modified to have  off-diagonal terms of $F$ set to zero.  The same model and test statistics are applied, but now there is truly no statistical dependence.   Fig. \ref{fig:4} shows the 2D SDE model (assuming $H = \frac{1}{2}$) achieves the expected rejection rates for the null hypothesis  (the GoF tests still detect artifacts of $H\ne \frac{1}{2}$).

Figure \ref{fig:4b} provides another illustration of robustness.  In this situation, the DGP process is once again the correlated
2D standard diffusion studied in Sec. \ref{sec:control}, but the fitted 2D SDE model constrains $R$ to zero (the other parameters
are fit via MLE).  Recall that in the DGP, the diagonal terms in $R$ are greater than zero (hence measurement noise is known to be present). 
The GoF tests readily detect the modeling error induced by setting $R$ equal to zero for all $N$ considered.  The independence tests reject relatively close to 
the nominal null level since statistical dependence induced by $F$ is accounted for explicitly.  However,   
measurement noise 
that is not properly modeled can cause non-Markovian effects. For example, increments of observations are anti-correlated in many situations relevant to SPT analysis
\cite{TSRV,Berglund2010,Weber2012,Ait-Sahalia2012}.  The coupling induced by non-zero off diagonal  components in $F$ of the 
DGP causes the normalized innovations (computed using the fitted model obtained where $R$ is set to zero) to become statistically dependent and this effect just starts 
to become detectable for the larger $N$ studied.  Making $F$ diagonal removes this effect (results not shown, but are similar to those shown in Fig. 
\ref{fig:4}).  Note that we took the extreme case of setting $R=0$, so other experimental artifacts  affecting the localization noise distribution, e.g. \cite{Thompson2002,Michalet2010a,Berglund2010,Enderlein2006,Gahlmann2013,Calderon2013b}, should also be readily handled by the independence testing approach discussed in this work.

\begin{figure}[htb]
\center
\centering

\begin{minipage}[b]{.7\linewidth}
\def\pw{1}

\begin{overpic}[width=\pw\textwidth]{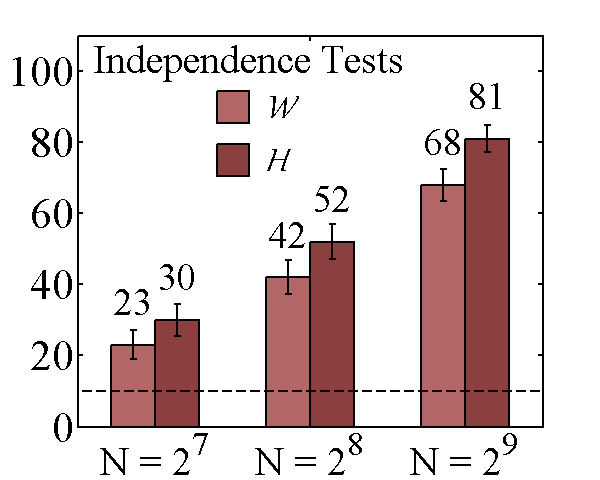}
% \put(15,72){\Large \color{black}(d)}
% \put(35,-5){{\makebox[1.\width]{\large 1D Model} \hfill }}
\put(-4,27){\rotatebox{90} {\makebox[1.\width]{\large $\%$ Rejected} \hfill }}
\put(12,-16){\mbox{\parbox[b]{0.8\textwidth}{\large Incorrect Zero Drift Model   \hfill } }}
% \put(50,-3){\mbox{\parbox[b]{0.5\textwidth}{\normalsize 1D Model [$H=\frac{1}{2}$] \hfill } }}
\end{overpic}
  \end{minipage}

\centerline{\footnotesize }
\caption{%\footnotesize 
\footnotesize 
(Color online).  Same as Fig. \ref{fig:1} except the assumed  model sets all drift parameters equal to zero.   Note: $\mathcal{W}= 33,54,77\%$ and $\mathcal{H}= 50,72,88\%$    for the 1D Model studied in Fig. \ref{fig:1}.
}
\label{fig:6}
\end{figure}

\begin{figure}[htb]
\center
\centering
\begin{minipage}[b]{1\linewidth}
\def\pw{.7}
\begin{overpic}[width=\pw\textwidth]{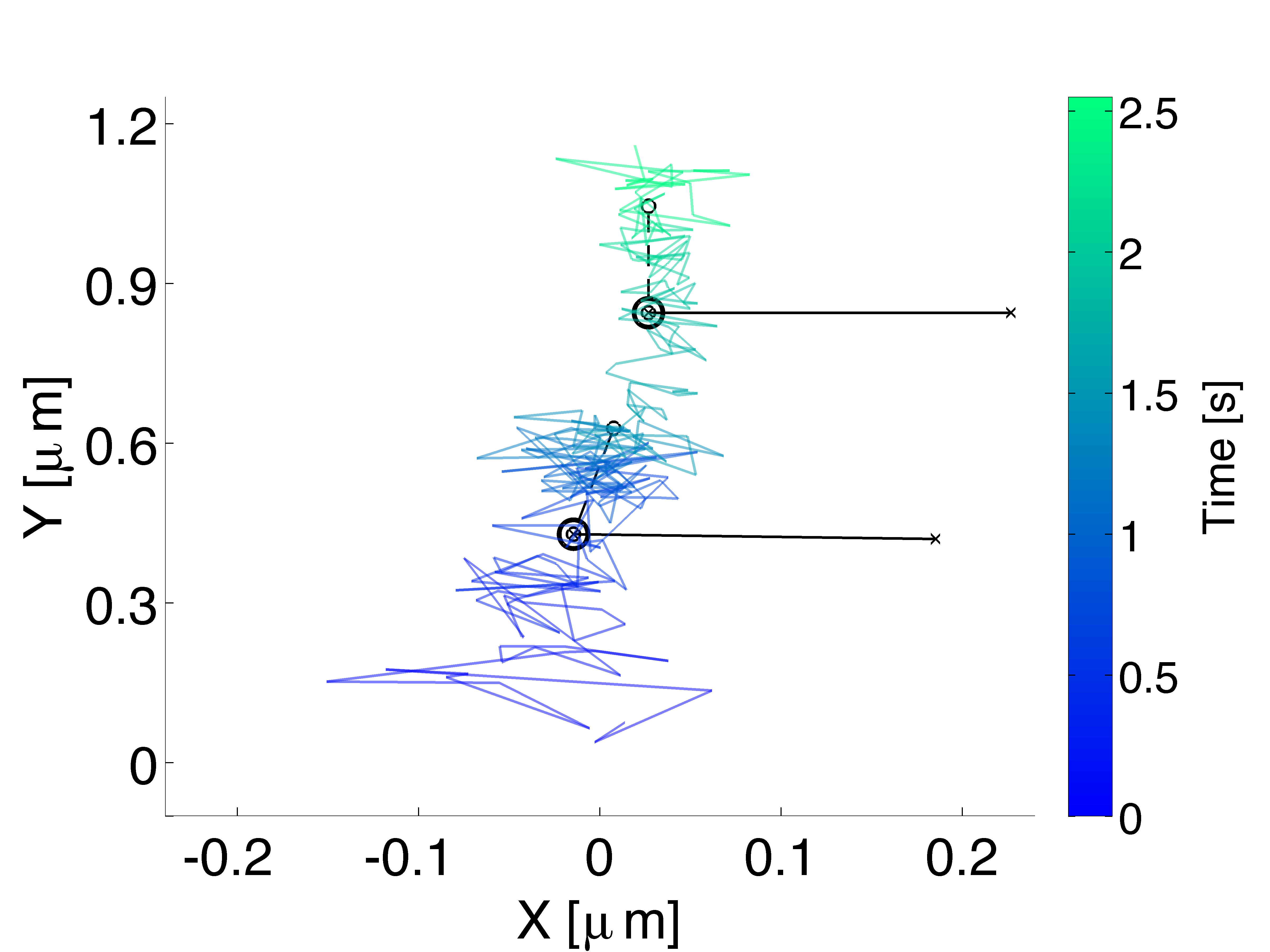} 
% \put(15,72){\Large \color{black}(a)}
% \put(-8,12){\rotatebox{90} {\makebox[1.\width]{\normalsize $\%$ Rejected} \hfill }}
% \put(0,-10){\mbox{\parbox[b]{0.45\textwidth}{\normalsize 2D SDE { \footnotesize [$H=0.5$ Model]} \hfill} }}
\end{overpic}  
\begin{overpic}[width=\pw\textwidth]{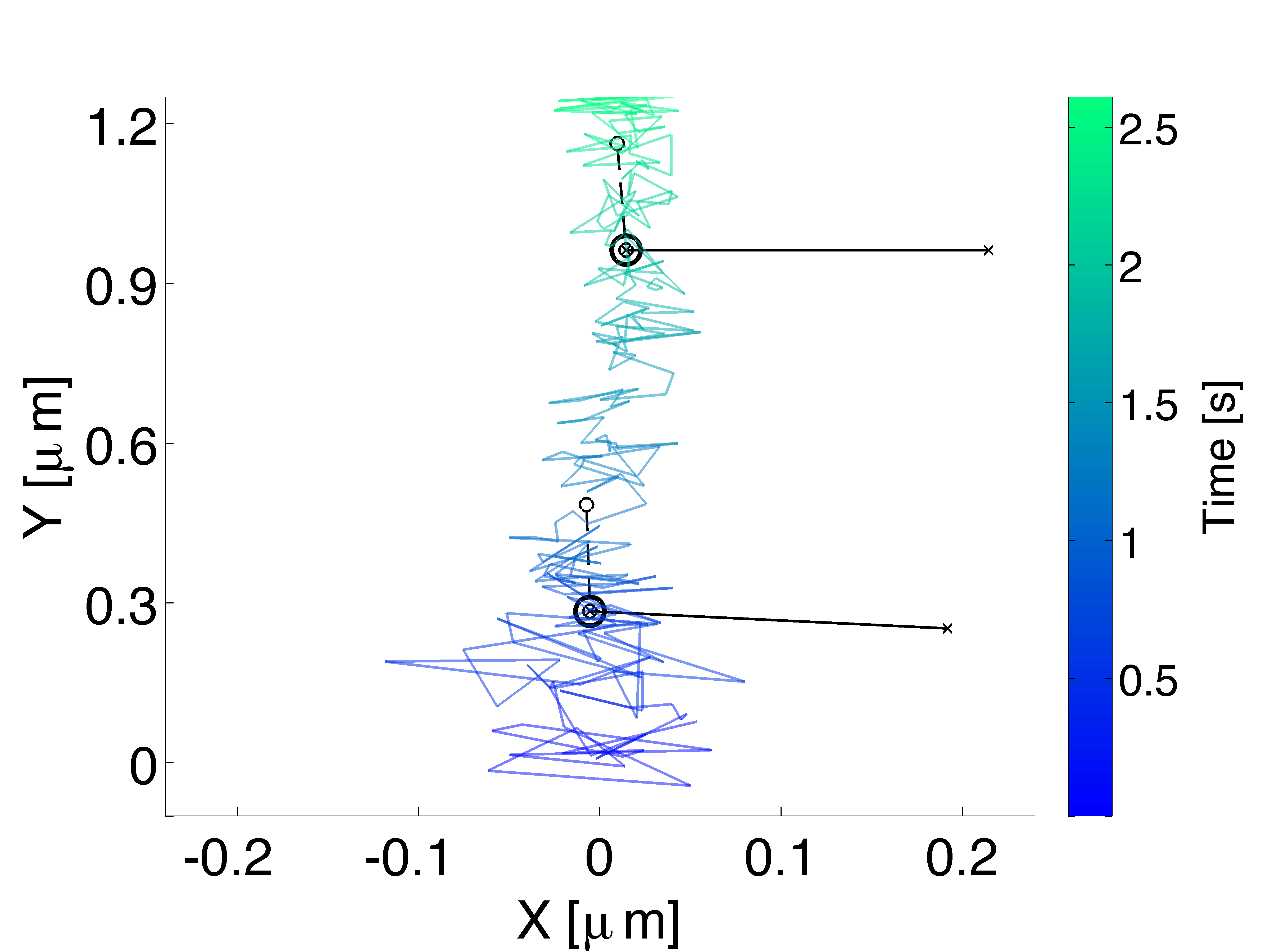}
% \put(15,72){\Large \color{black}(b)}
% \put(-8,12){\rotatebox{90} {\makebox[1.\width]{\normalsize $\%$ Rejected} \hfill }}
% \put(15,-5){{\makebox[1.\width]{\large 2D SDE ($H=0.5$) Model} \hfill }}
% \put(0,-10){\mbox{\parbox[b]{0.45\textwidth}{\normalsize 2D SDE { \footnotesize [$H=0.5$ Model]} \hfill} }}
\end{overpic}
  \end{minipage}

\centerline{\footnotesize }
\caption{%\footnotesize 
\footnotesize 
(Color online). YFP tagged IFT88 trajectories in the primary cilium of live epithelial cells.  Top: case where rejection caused by curved track occurred with $N=2^8$ points ($p$-vals of 2D and 1D models $<$ 0.05). Bottom: case where straight line motion was not rejected with $N=2^8$.  The arrows plotted denote the estimated eigenvectors of $F$ obtained by dividing the data into two disjoint time windows of size $N=2^7$ (eigenvectors corresponding to larger magnitude eigenvalue denoted by dark solid lines).  }
\label{fig:IFT}
\end{figure}

\subsection{Illustration of How the Assumed Stochastic Model Affects Statistical Power}

Although we just demonstrated robustness, this does not imply that
all models detect dependence between $x$ and $y$ with equal power. Fig. \ref{fig:6} uses the same DGP as in  \ref{fig:1}, but fits a models where $F$ is set to zero and all other parameters are estimated (i.e., the ``Zero Drift Model'').  Neglecting to account for the  linear autocorrelation induced by diagonal components of $F$ substantially reduces the power of $\mathcal{W}_N$ and $\mathcal{H}_N$; the unmodeled autocorrelation dominates the signal in the empirical marginals and complicates detecting  dependence between $x$ and $y$.  %Based on this observation, 
Hence, when aiming to detect statistical dependence, we suggest using estimation schemes advocated in Ref. \cite{Calderon2013b} since these models nest many classic SPT models as special cases.

\subsection{Application to Experimental IFT88 Trajectories}

Finally, we apply the techniques to analyze the IFT88 data described in the Methods section.  Fig. \ref{fig:IFT} plots two trajectories 
%(containing localization error and inherent thermal noise) 
where the molecular motors on IFT88  move cargo through the primary cilium \cite{Ye2013}.  
The total number of data points plotted corresponds to $N=2^8$;  in the top panel, the cytoskeletal track is curved and both  1D and 2D models assuming a constant $F$ are rejected using $\mathcal{H}_N$ with $N=2^8$.  In the bottom panel, 
the directed motion is effectively along a ``straight line" and the constant $F$ 2D model is not rejected by  
$\mathcal{H}_N$ or $\mathcal{W}_N$ with $N=2^8$.  We subsequently divided the trajectories into two disjoint time series of size $N=2^7$ and extracted the MLE of the 2D models;
the arrows in the plots show the eigenvectors corresponding to  
the estimated $F$ (the eigenvectors show the natural local coordinates).  
For the $N=2^7$, no tests applied resulted in rejection. The fact that the estimated eigenvector ``align'' with the direction of motion also suggests the $N=2^7$ is adequate to provide a local linear approximation of the curved cytoskeletal track.  These trajectories provide a clean graphical
example of nontrivial statistical dependence between measured $x,y$ positions. The technique presented detected unmodeled dependence in a nonstationary signal containing thermal and measurement noise, however, other nonlinear dependencies (beyond curvature) can be detected with the approach advocated in Ref. \cite{Duchesne2012}. 

\section{Conclusions} Hypothesis tests for detecting statistical  dependence between measured $x,y$ ($z$) data were demonstrated on simulated and experimental SPT  data.
 The approach was shown to be robust to controversial %modeling 
 assumptions (e.g., anomalous vs. normal diffusion) and exhibited reasonable power  when presented single trajectories with small $N$ and non-negligible measurement noise using parameters characteristic of SPT experiments.  
The measurement noise issue is particularly relevant to SPT applications since complex background and other optics effects can complicate reliably modeling and extracting parameters characterizing  noise induced by the measurement apparatus \cite{Thompson2002,Michalet2010a,Berglund2010,Enderlein2006,Gahlmann2013,Calderon2013b}. 
Such effects often become more pronounced as the temporal resolution 
increases \cite{Ait-Sahalia2012}, so the ``robustness'' aspect
of the testing procedure  shown is particularly appealing from an applied experimental data analysis perspective. 
% Recall we took the extreme position of completely ignoring the measurement noise but modeled statistical dependence;  the GoF tests readily identified this error, but the independence tests 

\section{Acknowledgments}
The authors thank  L. Milenkovic and M.P. Scott for experimental help and K. Ghoudi for sharing code illustrating
an efficient computational implementation  of the 3D method presented in Ref. \cite{Duchesne2012}.
 CPC's work was supported by NSF SBIR Grant No. 1314897; LEW \& WEM's by NIH Grant  No. R01GM086196 and a Stanford Bio-X Graduate Student Fellowship.

\section{Appendix}
The M\"{o}bius transformation for the test of primary interest in our study is defined by:

\begin{align}
\label{eq:mob}
 \mathbb{R}_{A,\vec{\ell},N}(\vec{x})= & N^{-\frac{1}{2}}\sum \limits_{k=1}^N \prod \limits_{j \in A} [\mathbb{I}\{ e_{j, k+l_j } \le x_j \} - F_{jN}(x_j) ], \\
% R_{j,t+l_j}\equiv & F_{jN}( e_{j, t+l_j }):=  \frac{1}{N}\sum \limits_{k=1}^N 
 & F_{jN}( e_{j, t+l_j }):=   \frac{1}{N}\sum \limits_{k=1}^N
\mathbb{I}\{ e_{jk} \le  e_{j, k+l_j }\}.
\end{align}

In the expression above, $\mathbb{I}\{\cdot\}$ represents the indicator function of an event;  $\vec{\ell}$  specifies a temporal offset (or ``time lag" \cite{Duchesne2012}) vector;
$x_j$ ($l_j$) is component $j$ of $\vec{x} \in \mathbb{R}^d$ ($\vec{\ell} \in \mathbb{R}^d$); and $F_{jN}$ denotes the empirical marginal distribution   of residual $j$ estimated from $N$ observations \cite{Duchesne2012}. The set $A$ is a collection of position indices (e.g., $A=(1,3)$ represents the $x$ and $z$ coordinates).  
Let $\mathcal{A}$ denote
the collection of all possible spatial interactions, e.g. if $d=3$ then  
$\mathcal{A}=\{(1,2,3),(1,2),(1,3),(2,3) \}$.  
Despite the fact that  $\mathbb{R}_{A,\vec{\ell},N}$ and $\mathbb{R}_{B,\vec{\ell},N}$ for $A,B\in \mathcal{A}$ can have overlapping empirical rank information,   the M\"{o}bius transformation (Eq. \ref{eq:mob})  has the  property that 
$\mathrm{cov}(\mathbb{R}_{A,\vec{\ell},N},\mathbb{R}_{B,\vec{\ell}^\prime,N})=0$  unless $A=B$ and $\vec{\ell} = \vec{\ell}^\prime$ in the limit $N\rightarrow \infty$ (additional transformations can produce standard Gaussian vectors \cite{Duchesne2012}); these fact can be exploited to construct test statistics generated by combining multiple lag vectors ($\vec{\ell}$) and position index sets ($A$) into a single test statistic. 

As mentioned in the main text, we are primarily interested in the generalized cross-correlations.  
This quantity is defined by:

\begin{align}
\label{eq:gencov}
 \nonumber N^{\frac{1}{2}} \hat{\gamma}_{A,\vec{\ell},N} & = (-1)^{|A|} \int\limits_{\mathbb{R}^{|A|}}
 \mathbb{R}_{A,\vec{\ell},N}(\vec{x}^{(A)})d\vec{x}^{(A)} \\
  = & N^{-\frac{1}{2}}\sum \limits_{k=1}^N \prod \limits_{j \in A} (e_{j, k+l_j }   - \bar{e}_j)  
\end{align}

where  $|A|$ is the number of elements in $A$, $\vec{x}^{(A)}$ are the coordinates of $\vec{x}$ contained in $A$, 
and
$\bar{e}_j := \frac{1}{N}\sum \limits_{k=1}^N e_{j, k } $
The second line above follows from the definition of $\mathbb{R}_{A,\vec{\ell},N}$ and the fact that  the mean of a continuous 
random variable $X$ (having distribution function $F(t):=P(X \le t)$)
is given by  
$\mathbb{E}[X] = \int \limits_{0}^\infty \big(1-F(t)\big)dt - \int \limits_{-\infty}^0  F(t)dt$ for real-valued random variables \cite{Cinlar1997} (the multivariate analog of this identity is used above \cite{Duchesne2012}).     In our work, we use the same time lag parameter vectors utilized in Ref. \cite{Duchesne2012}  throughout.
The primary test statistic of interest, $\mathcal{H}_N$, is readily computed using these expressions (the relevant expressions for $\mathcal{W}_N$ are provided in Ref. \cite{Duchesne2012}).

\bibliographystyle{nature}
\bibliography{running,StatPapers,biomedtracking,tracking,Primary_Cilium,mRNAII,mRNA,HIV_Gag,Primary_Cilium_Shared}

\begin{thebibliography}{10}

\bibitem{Arhel2006}
Arhel, N., Genovesio, A., Kim, K., Miko, S., Perret, E., Olivo-Marin, J.,
  Shorte, S., and Charneau, P.
\newblock {\em Nat. Methods}{ \bf 3}(10), 817--824 (2006).

\bibitem{Golding2006a}
Golding, I. and Cox, E.
\newblock {\em Phys. Rev. Lett.}{ \bf 96}(9), 14--17 March  (2006).

\bibitem{Lessard2007}
Lessard, G.~A., Goodwin, P.~M., and Werner, J.~H.
\newblock {\em Appl. Phys. Lett.}{ \bf 91}(22), 224106 November  (2007).

\bibitem{Lange2008}
Lange, S., Katayama, Y., Schmid, M., Burkacky, O., Br\"{a}uchle, C., Lamb,
  D.~C., and Jansen, R.-P.
\newblock {\em Traffic}{ \bf 9}(8), 1256--67 August  (2008).

\bibitem{manley2008}
Manley, S., Gillette, J., Patterson, G., Shroff, H., Hess, H., Betzig, E., and
  Lippincott-Schwartz, J.
\newblock {\em Nat. Methods}{ \bf 5}(2), 155--157 (2008).

\bibitem{Thompson2010}
Thompson, M.~A., Casolari, J.~M., Badieirostami, M., Brown, P.~O., and Moerner,
  W.~E.
\newblock {\em Proc. Natl. Acad. Sci. U. S. A.}{ \bf 107}(42), 17864--71
  October  (2010).

\bibitem{Weigel2011}
Weigel, A.~V., Simon, B., Tamkun, M.~M., and Krapf, D.
\newblock {\em Proc. Natl. Acad. Sci. U. S. A.}{ \bf 108}(16), 6438--43 April
  (2011).

\bibitem{Ram2012}
Ram, S., Kim, D., Ober, R.~J., and Ward, E.~S.
\newblock {\em Biophys. J.}{ \bf 103}(7), 1594--603 October  (2012).

\bibitem{Ye2013}
Ye, F., Breslow, D.~K., Koslover, E.~F., Spakowitz, A.~J., Nelson, W.~J., and
  Nachury, M.~V.
\newblock {\em Elife}{ \bf 2}, e00654 (2013).

\bibitem{Verdaasdonk2013}
Verdaasdonk, J.~S., Vasquez, P.~A., Barry, R.~M., Barry, T., Goodwin, S.,
  Forest, M.~G., and Bloom, K.
\newblock {\em Mol. Cell}{ \bf } (2013).

\bibitem{Stephens2013}
Stephens, A.~D., Snider, C.~E., Haase, J., Haggerty, R.~a., Vasquez, P.~a.,
  Forest, M.~G., and Bloom, K.
\newblock {\em J. Cell Biol.}{ \bf 203}(3), 407--16 November  (2013).

\bibitem{Masson2009}
Masson, J., Casanova, D., Turkcan, S., Voisinne, G., Popoff, M., Vergassola,
  M., and Alexandrou, A.
\newblock {\em Phys. Rev. Lett.}{ \bf 102}(4), 48103 (2009).

\bibitem{Voisinne2010}
Voisinne, G., Alexandrou, A., and Masson, J.-B.
\newblock {\em Biophys. J.}{ \bf 98}(4), 596--605 February  (2010).

\bibitem{Calderon2013b}
Calderon, C.~P., Thompson, M.~A., Casolari, J.~M., Paffenroth, R.~C., and
  Moerner, W.~E.
\newblock {\em J. Phys. Chem. B}{ \bf } October  (2013).

\bibitem{Kou2004}
Kou, S. and Xie, X.
\newblock {\em Phys. Rev. Lett.}{ \bf 93}(18), 180603 October  (2004).

\bibitem{kou_08}
Kou, S.
\newblock {\em Annals of Applied Statistics}{ \bf 2}, 501--535 (2008).

\bibitem{Magdziarz2010}
Magdziarz, M. and Klafter, J.
\newblock {\em Phys. Rev. E}{ \bf 82}(1), 011129 July  (2010).

\bibitem{Weber2012}
Weber, S.~C., Thompson, M.~A., Moerner, W.~E., Spakowitz, A.~J., and Theriot,
  J.~A.
\newblock {\em Biophys. J.}{ \bf 102}(11), 2443--50 June  (2012).

\bibitem{Kepten2013}
Kepten, E., Bronshtein, I., and Garini, Y.
\newblock {\em Phys. Rev. E}{ \bf 87}(5), 052713 May  (2013).

\bibitem{Meroz2013}
Meroz, Y., Sokolov, I.~M., and Klafter, J.
\newblock {\em Phys. Rev. Lett.}{ \bf 110}(9), 090601 February  (2013).

\bibitem{Bickel2006}
Bickel, P.~J., Ritov, Y., and Stoker, T.~M.
\newblock {\em Ann. Stat.}{ \bf 34}(2), 721--741 April  (2006).

\bibitem{Berglund2010}
Berglund, A.~J.
\newblock {\em Phys. Rev. E}{ \bf 82}(1), 011917 July  (2010).

\bibitem{Ait-Sahalia2010}
A\"{\i}t-Sahalia, Y., Fan, J., and Xiu, D.
\newblock {\em J. Am. Stat. Assoc.}{ \bf 105}(492), 1504--1517 December
  (2010).

\bibitem{Michalet2012}
Michalet, X. and Berglund, A.
\newblock {\em Phys. Rev. E}{ \bf 85}(6), 061916 June  (2012).

\bibitem{TSRV}
Zhang, L., Mykland, P., and Ait-Sahalia, Y.
\newblock {\em Journal of the American Statistical Association}{ \bf 100},
  1394--1411 (2005).

\bibitem{Ait-Sahalia2012}
A\"{\i}t-Sahalia, Y. and Jacod, J.
\newblock {\em J. Econ. Lit.}{ \bf 50}(4), 1007--1050 December  (2012).

\bibitem{Calderon2013}
Calderon, C.~P.
\newblock {\em Phys. Rev. E}{ \bf 88}(1), 012707 April  (2013).

\bibitem{Duchesne2012}
Duchesne, P., Ghoudi, K., and Remillard, B.
\newblock {\em Can. J. Stat.}{ \bf 40}(3), 447--479 (2012).

\bibitem{Note1}
The drift function in the $H=\protect \frac {1}{2}$ case is motivated by the
  overdamped Langevin equation; for $H\not =\protect \frac {1}{2}$, a memory
  kernel is not used; however our interest is in robustly detecting statistical
  dependence between components regardless of fluctuation dissipation
  constraints.

\bibitem{hong}
Hong, Y. and Li, H.
\newblock {\em Rev. Fin. Studies}{ \bf 18}, 37--84 (2005).

\bibitem{Tran2008}
Tran, P.~V., Haycraft, C.~J., Besschetnova, T.~Y., Turbe-Doan, A., Stottmann,
  R.~W., Herron, B.~J., Chesebro, A.~L., Qiu, H., Scherz, P.~J., Shah, J.~V.,
  Yoder, B.~K., and Beier, D.~R.
\newblock {\em Nat. Genet.}{ \bf 40}(4), 403--10 April  (2008).

\bibitem{Ott2012}
Ott, C. and Lippincott-Schwartz, J.
\newblock {\em Curr. Protoc. Cell Biol.}{ \bf Chapter 4}(December), Unit 4.26
  December  (2012).

\bibitem{Thompson2002}
Thompson, R.~E., Larson, D.~R., and Webb, W.~W.
\newblock {\em Biophys. J.}{ \bf 82}(5), 2775--83 May  (2002).

\bibitem{Michalet2010a}
Michalet, X.
\newblock {\em Phys. Rev. E}{ \bf 82}(4), 041914 October  (2010).

\bibitem{Enderlein2006}
Enderlein, J., Toprak, E., and Selvin, P.~R.
\newblock {\em Opt. Express}{ \bf 14}(18), 8111--20 September  (2006).

\bibitem{Gahlmann2013}
Gahlmann, A. and Moerner, W.~E.
\newblock {\em Nat. Rev. Microbiol.}{ \bf 12}(1), 9--22 December  (2014).

\bibitem{Cinlar1997}
Cinlar, E.
\newblock {\em {Introduction to Stochastic Processes}}.
\newblock Prentice Hall College Div,  (1997).

\end{thebibliography}

\end{document}